\documentclass[12pt]{article}
\usepackage[T1]{fontenc}
\usepackage[latin9]{inputenc}
\usepackage[letterpaper]{geometry}
\usepackage{amsmath}
\usepackage{amsthm}
\usepackage{amssymb}
\usepackage{subfig}
\usepackage{float}
\usepackage{graphicx,psfrag,epsf}
\usepackage{enumerate}
\usepackage{natbib}
\usepackage{hyperref}

\begin{document}
\def\spacingset#1{\renewcommand{\baselinestretch}%
{#1}\small\normalsize} \spacingset{1}

\title{Original Research Paper\vspace{.2in}\\ A Min/Max Algorithm for Spline Based Modeling of Violent Crime Rates in USA\footnote{Short Running Title: A Min/Max Algorithm for Spline Modeling}}
\author{Eric Golinko and Lianfen Qian\footnote{Corresponding author: Lianfen Qian is funded by OURI Curriculum Grant from Florida Atlantic University and ZJNSF LY17A010012 as a specially appointed professor of Wenzhou University, Zhejiang, China. 
Her contact email address is: Lqian@fau.edu, 001-561-297-2486, fax: 001-561-297-2436} }

\date{Florida Atlantic University, Boca Raton, FL 33431}
\maketitle

\begin{abstract}  This paper focuses on modeling violent crime rates against population over the years 1960-2014 for the United States via cubic spline based method.  We propose a new min/max algorithm on knots detection and estimation for cubic spline regression. We employ least squares estimation to find potential regression coefficients based upon the cubic spline model and the knots chosen by the min/max algorithm. We then utilize the best subsets regression method to aid in model selection in which we find the minimum value of the Bayesian Information Criteria. Finally, we report the $R_{adj}^{2}$
  as a measure of overall goodness-of-fit of our selected model. Among the fifty states and Washington D.C., we have found 42 out of 51 with $R_{adj}^{2}$
  value that was greater than $90\%$. We also present an overall model for the United States as a whole.  Our method can serve as a unified model for violent crime rate over future years.\\
  
\end{abstract}

\section{Introduction}
Violent crimes as defined by the US Department of Justice are offenses classified as murder, robbery, rape, and aggravated assault.  It is of interest to investigate the interaction of violent crimes in relation to the ever growing and diverse population. As indicated in Kurtz et al. (2007),
the United States population is growing according to an exponential model, the natural research topics are to explore how does population growth affect violent crime rate and what is an appropriate technique to show this relationship. In this paper, we investigate violent crime rate and population data across fifty years for all states in United States, as well as the District of Columbia. We develop a min/max algorithm which employs the cubic spline regression model with a new automatic knots selection procedure to model the violent crime rate trend against the time and population size. The population size is estimated by a  recursive exponential model in Robinson et al. (1993).


In spline modeling, one of the crucial key is the knot selection. Under normal error assumption,  Molinari et al. (2004) 
presents an algorithm on finding optimal knots for regression splines using AIC criteria for least squares estimates. Spiriti et al. (2013)
offers a different algorithm on knots selection with penalized splines. These papers focus on computation complexity with applications in social science and computational statistics.  In this paper, we propose a simple automatic knot selection algorithm as an alternative to overcome the computational complexity.

In this paper, we gather violent crime and population size for the fifty states and Washington District of Columbia (for short, D.C. and also will be treated as a state) in USA over the years 1960-2014. The  population size is from the United States Census Bureau. Notice that U.S. Census takes place every ten years. Hence the population size consists of  the actual census population size for the years of (1960, 1970, 1980, 1990, 2000 and 2010) and estimates by Kurtz et al. (2007)
for intercensal years. The violent crime frequency is collected via a customizable table tool available at \url{https://www.ucrdatatool.gov/} from the US Department of Justice.  The data set is organized on a state wide basis, including D.C., over the years of  1960-2014.

 The objective of the paper is to find a unified model in which we can show the relationship of violent crime and population over this time period for all states, as well as the U.S. as a whole. This model is intended only to rely on the information of a given state's population and a respective rate of violent crime. We use the technique of cubic spline regression, coupled with a new min/max algorithm for automatic knot selection that defines a partition of the data and  finds absolute maximum deviates from a partition mean. We find in this approach a pragmatic solution to which all a model requires is a partition for a given set of data. The approach returns significant coefficients of cubic spline model with automatically detected knots by the proposed min/max algorithm.

\section{Exploratory Data Analysis of Violent Crime Rate}

In our preliminary analysis we note that New York did not have violent crime counts from 1960-1964, however this did not impact our overall process, rather we were able to formulate the same results for New York by simply analyzing the data that was available for 1965-2014. In all other states we look at the full range of years from 1960-2014. The totals for the years 1960-2014 of the entire USA were also collected in this way, as with table building tool we can specify to organize by state, or an overall national average. Specifically it should be noted that these years were selected specifically because the Department of Justice table building tool offered these years for all states in this time range, but not preceding. It is also thought that a fifty year context was sufficient to shape a model for all states and Washington D.C. A simple violent crime rate is defined as below:

\begin{equation}
VCR_t(State)=\frac{\mbox{Violent~Crime~Frequency~of~State ~in~$i^{th}$~year}}{\mbox{Population~Size~of~State~in~$i^{th}$~year}}, \ i=1960,...,2014.
\label{eq:vcr}
\end{equation}

\begin{figure}
    \includegraphics[width=\columnwidth]{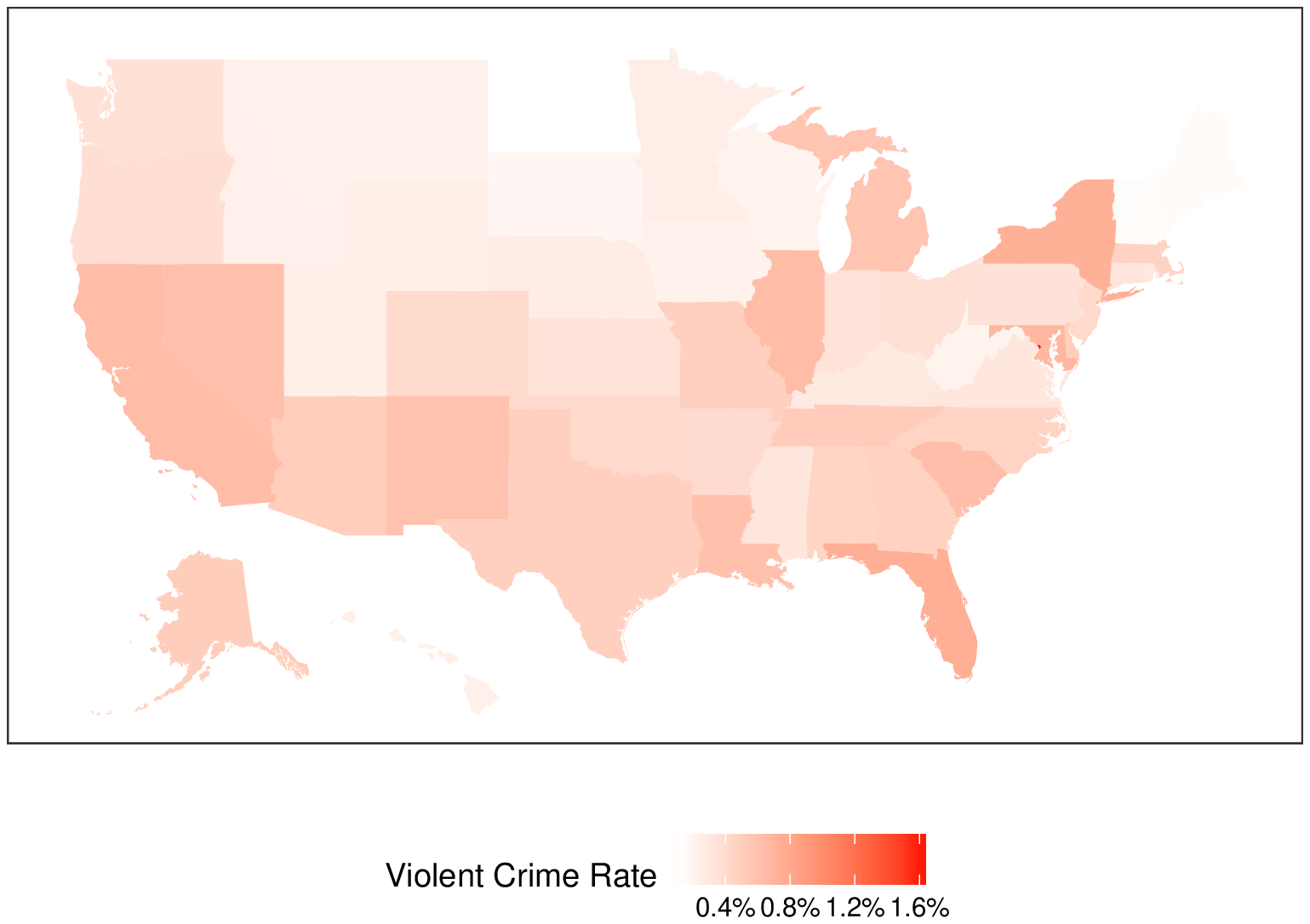}
    \caption{Average violent crime rate in USA from 1960 to 2014}
\label{fig:vc_rate}    
\end{figure}
\begin{figure}
    \centering
    \includegraphics[height=3in, width=\textwidth]{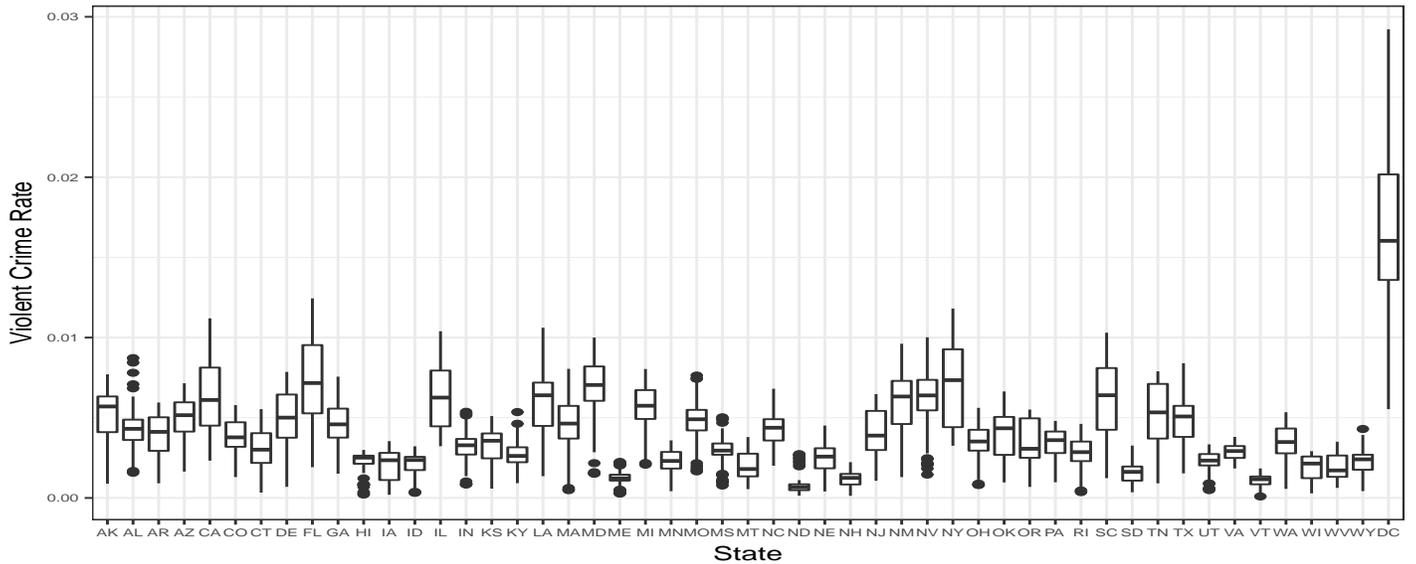}
    \caption{Boxplot of violent crime rate, 1960-2014}
    \label{fig:vcr_box}
\end{figure}

Figure~\ref{fig:vc_rate} shows the state by state heatmap of the violent crime rate
over the years 1960-2014. Here, we can observe the relationship of localities that contain large cities have a higher prevalence of violent crime  than  other states.

Figure~\ref{fig:vcr_box}  is  the boxplot of the violent crime rate for all states
over the years 1960-2014. Here, we notice that 
Washington D.C.'s violent crime rate is well above that of other
states. This may be due to the concentration of this area  as a city.  
Other major states in the US are surrounded by suburban and rural
areas, while Washington D.C. is predominantly an urban area. 

\begin{figure}[H]
\centering
 \subfloat[USA violent crime rate]
 {\includegraphics[width=0.5\textwidth]{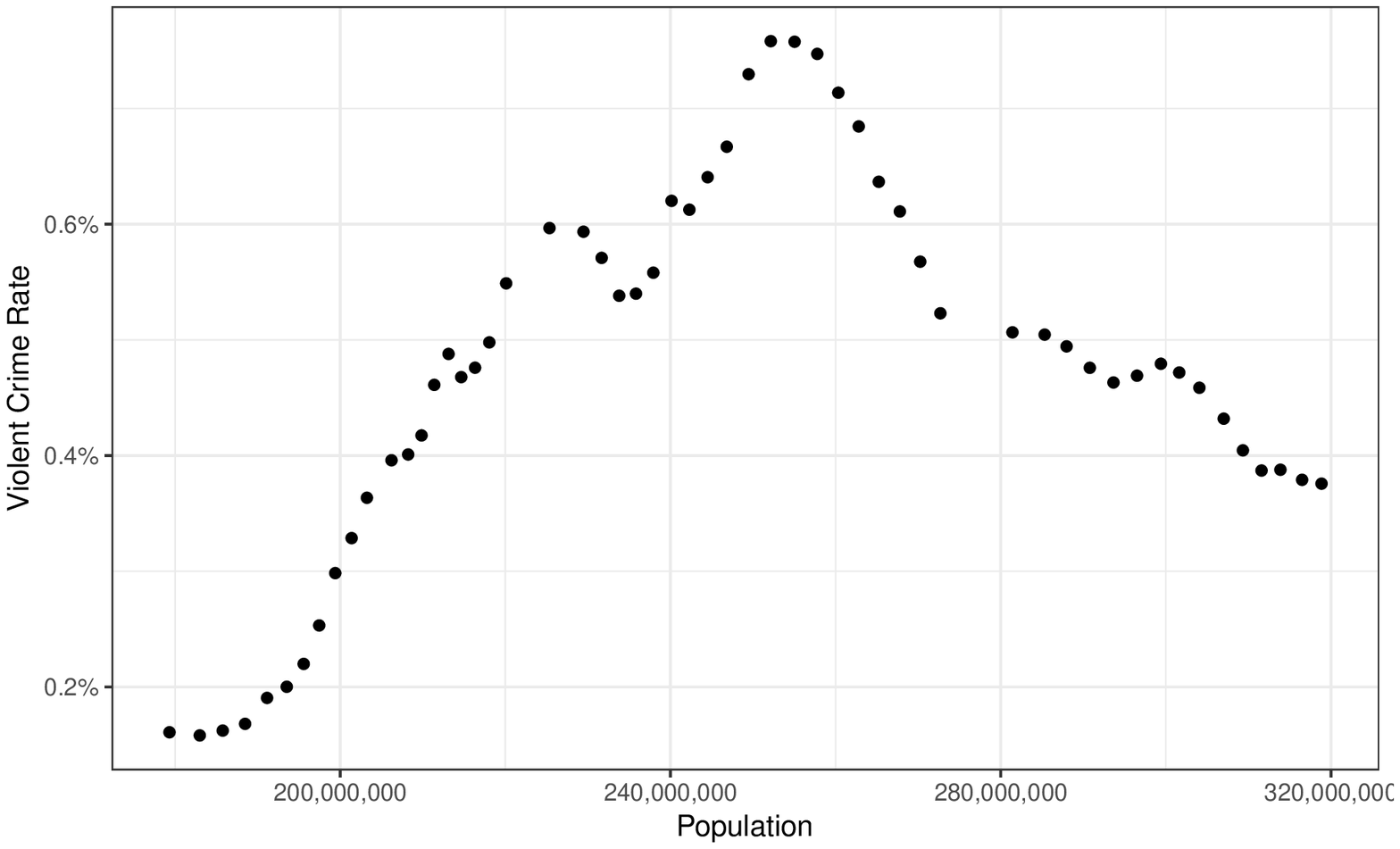}\label{fig:usa-vc-rate}}
 \hfill
 \subfloat[USA population]
 {\includegraphics[width=0.5\textwidth]{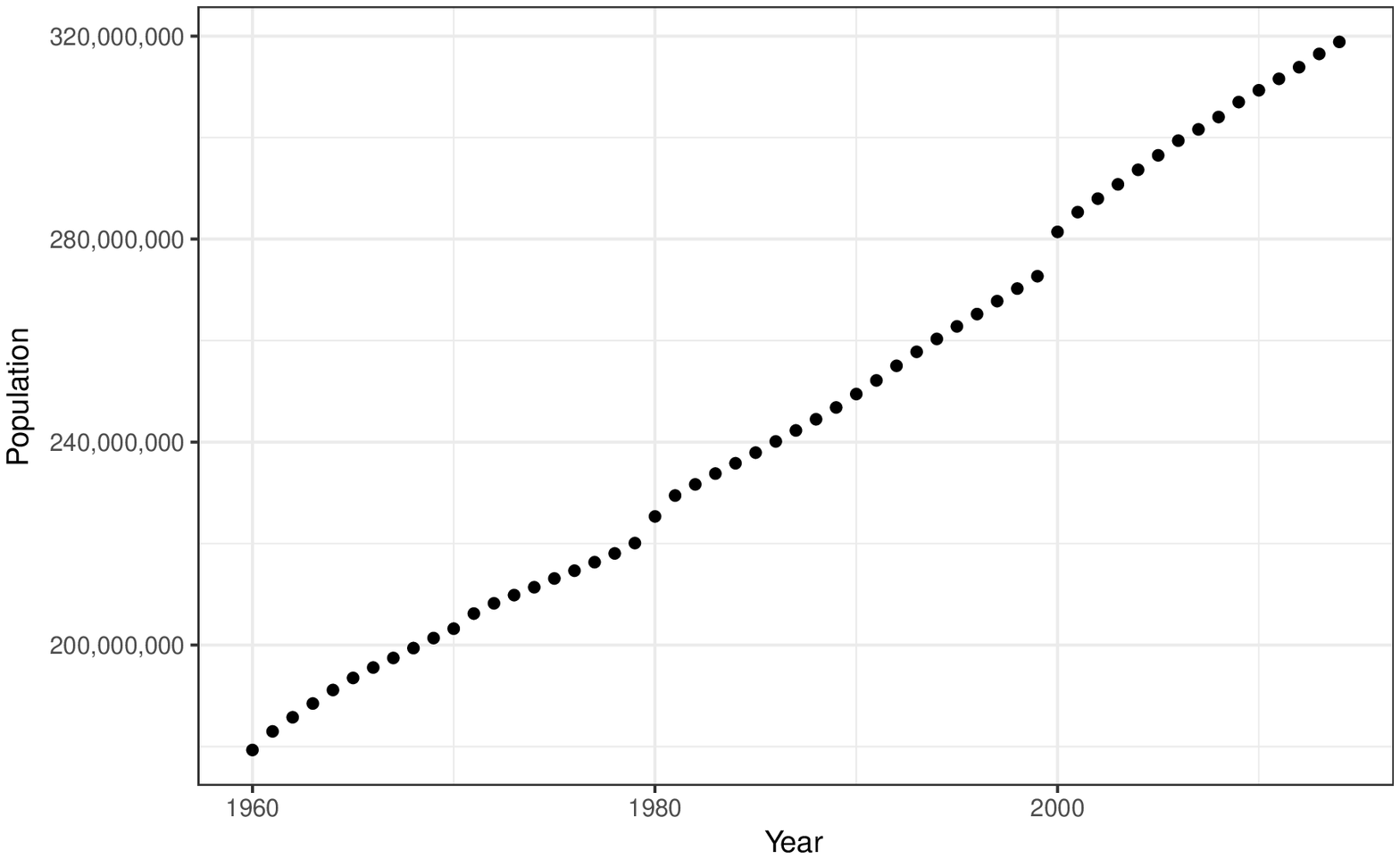}\label{fig:usa_rate}}
 \caption{USA violent crime rate and population}
\label{fig:usa_eda} 
\end{figure}

Figure~\ref{fig:usa_eda} shows the national non-linear trend of violent crime rate versus population size, as well as the linear increasing population trend over time.  We want to find an appropriate unified model to describe this non-linear relationship of violent crime rate and population both nationally and state-wise.  We note that the \textit{x-axis }is based upon population, if a state's population is increasing over time, graphically, if we were to replace the \textit{x-axis} by time the scatter plot would look the similar.

With these instances included in the data for all fifty states, we noticed three different patterns of population trend over time. Firstly, a linearly increasing trend mirrors the United States overall pattern where population increases over time such as in Figure 3(b) for Florida. Second and third patterns are non-linear increasing and decreasing with possible sudden drop and rise, respectively. Illustrations of states are shown in Figures ~\ref{fig:la_eda}-\ref{fig:dc_eda}. The irregular non-linear trend with potential sudden change  in population over time adds an additional challenge to find a unified model for all fifty states and Washington D.C.  

\begin{figure}[H]
\centering
  \subfloat[Florida violent crime rate]{\includegraphics[width=0.5\textwidth]{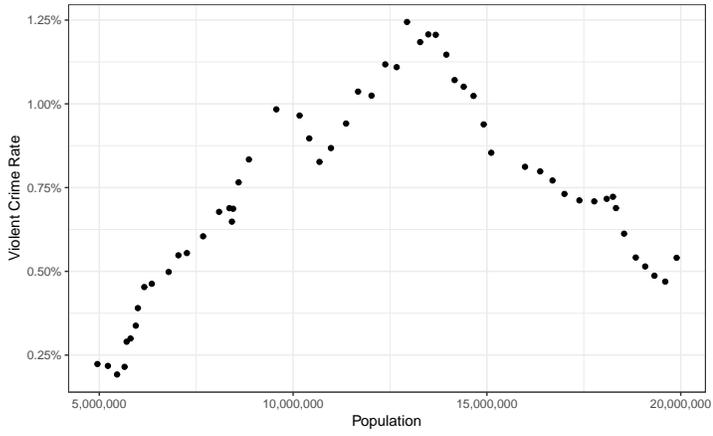}\label{fig:fl-vc-rate}}
  \hfill
  \subfloat[Florida population]{\includegraphics[width=0.5\textwidth]{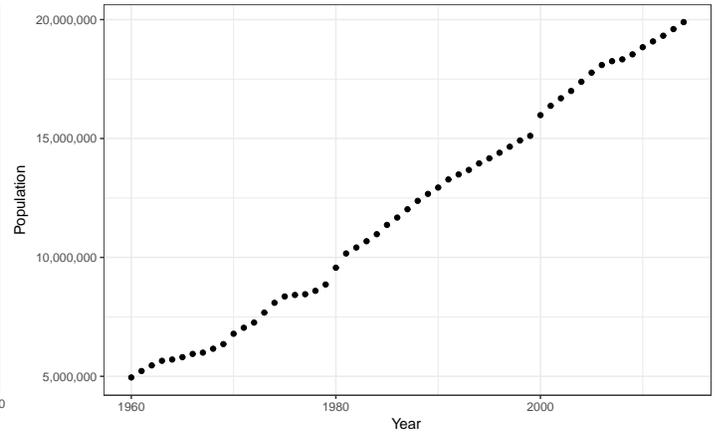}\label{fig:fl_rate}}
  \caption{Florida violent crime rate and population}
\label{fig:fl_eda} 
\end{figure}

\begin{figure}[H]
\centering
  \subfloat[Louisiana violent crime rate]{\includegraphics[width=0.5\textwidth]{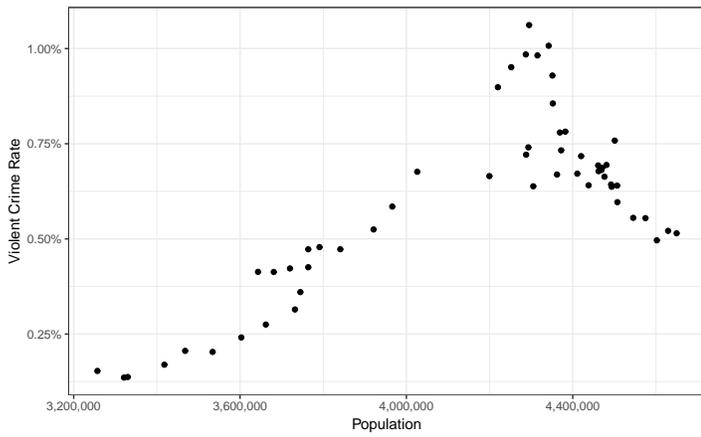}\label{fig:la-vc-rate}}
  \hfill
  \subfloat[Louisiana population]{\includegraphics[width=0.5\textwidth]{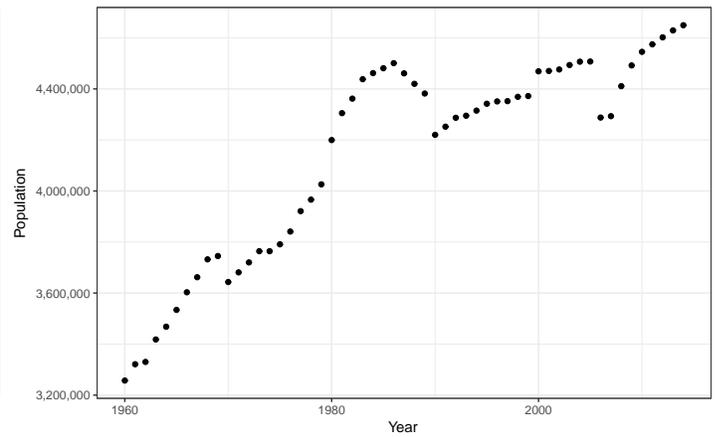}\label{fig:la_rate}}
  \caption{Louisiana violent crime rate and population}
\label{fig:la_eda} 
\end{figure}

\begin{figure}[H]
\centering
  \subfloat[Washington DC violent crime rate]{\includegraphics[width=0.5\textwidth]{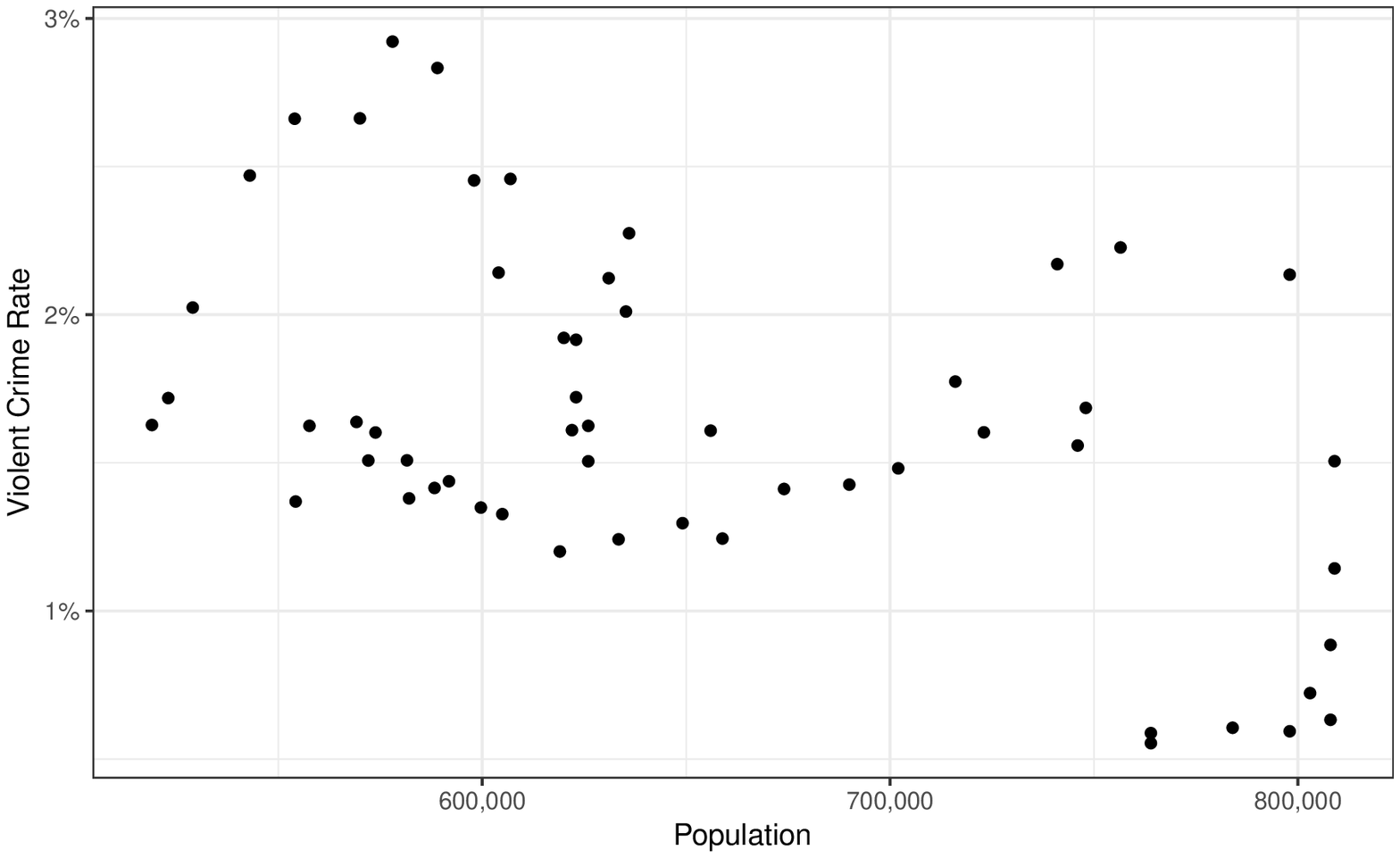}\label{fig:dc-vc-rate}}
  \hfill
  \subfloat[Washington DC population]{\includegraphics[width=0.5\textwidth]{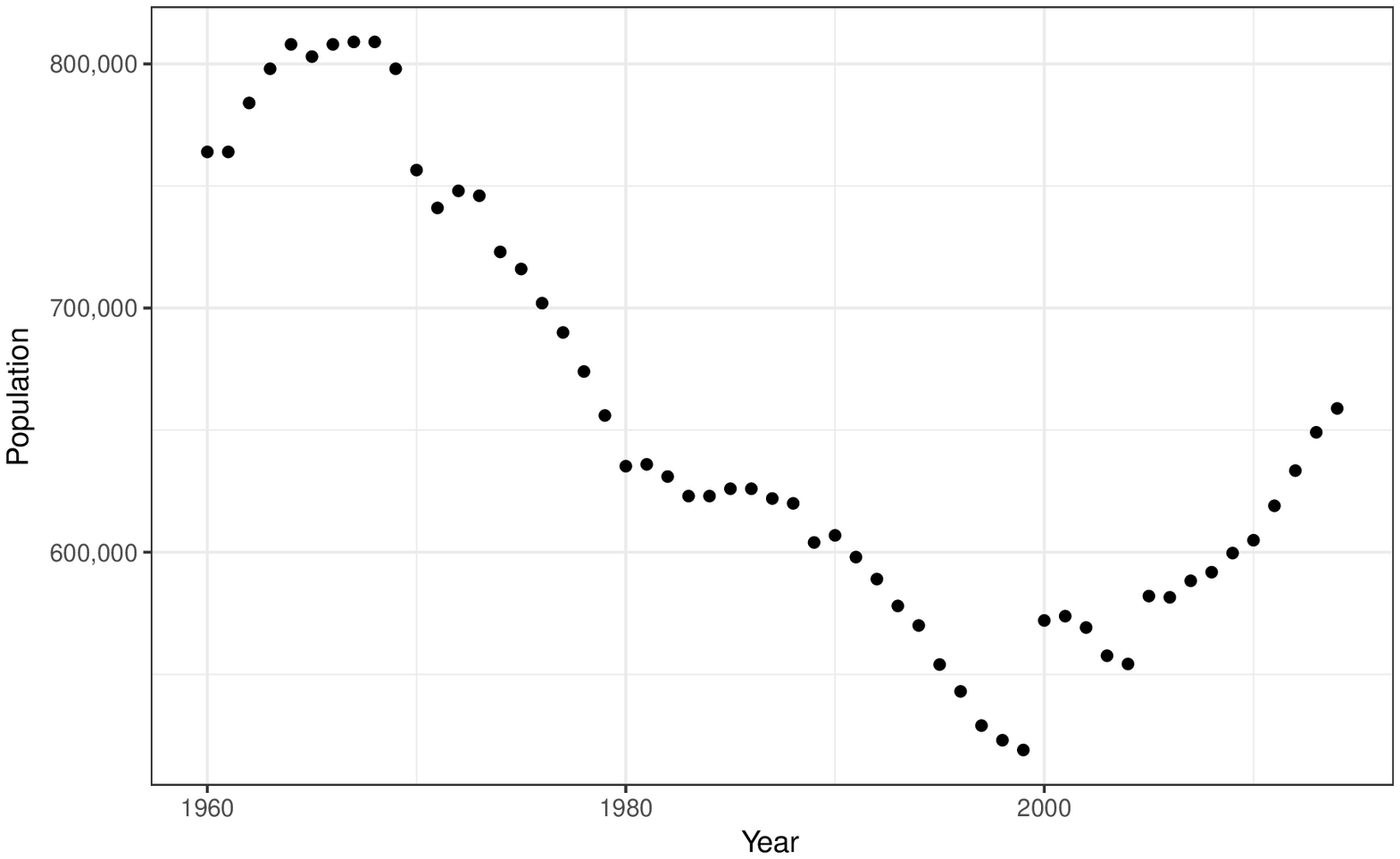}\label{fig:dc_rate}}
  \caption{Washington DC violent crime rate and population}
\label{fig:dc_eda} 
\end{figure}

\section{Cubic Spline Modeling with A Min/Max Algorithm} 

Let
$y_i$ be the violent crime rate   and $x_i$ be the population size in $i$th year of each state or USA. We define the general cubic spline model with $K$ knots based on a predictor $x_i$. We consider $n$ years of violent crime data. Let the $k^{th}$ knot location be $t_{0k}\in\{x_{1},..,x_{n}\}$ for $k=1,...,K$, where $x_i$ is the population size for $i$th year, $i=1,..., n$.  Then the cubic spline model with mean function as below: 
\begin{equation}
\begin{split}
E(y)=\beta_{0}+\beta_{1}x+\beta_{2}x^{2}+\beta_{3}x^{3}+\sum\limits _{k=1}^{K}\beta_{0k}(x-t_{0k})_{+}^{3}, ~~ where \\
(x-t_{0k})_{+}^{3}=
\begin{cases}
\begin{array}{cc}
0, & \mbox{if } x-t_{0k}\leq0,\\
(x-t_{0k})^{3}, & \mbox{if } x-t_{0k} >0.
\end{array}
\end{cases}
\end{split}
\label{eq:spline-def}
\end{equation}

With this definition, we can think of the cubic spline model as being a piecewise cubic regression model, such that for pre-selected knots
we either include values of the predictor at a knot $t_{0k}$, or else the value is zero.  We  also may add other predictors to the spline to further a multiple regression model.

To describe the proposed min/max \textit{K-part}  algorithm, we first pre-select $K$. If $K$ divides the number of observations, the data set is divided into $K$ parts. If $K$ does not divide $n$, that is $n\,mod\,K=r$, $r\neq0$, then our algorithm will still estimate potential knots over the first $K$ partitions, and the remaining values will be used only in our overall modeling process. From a practical perspective, this is a motivation for choosing a $K$ which divides $n$ if possible. In this section, we first  propose the new  $K$-partition algorithm and its implementation, then discuss the STATE model, which we use to model the fifty states, Washington D.C., and the entire USA. 

Let $L=\left[\frac{n}{K}\right]$ and $y_i$ be the violent crime rate in the $ith$ year for a state or USA.  We  begin to partition $(y_{1},...,y_{n})$ , such that the $k^{th}$ partition is $\{y_{(k-1)L+\ell},...,y_{kL}\},$ $1\leq\ell\leq L$ and $1\leq k\leq K$. Then we compute the $k^{th}$ partition sample mean. We designate this as: 
$$\bar{y}_{k}=\frac{\sum\limits _{\ell=1}^{L}y_{(k-1)L+\ell}}{L}, \ k=1,..,K.$$
Once we have computed the partition sample means, we evaluate the difference of each value of the partition with it's respective mean and pick the maximum: 
$$\max_{1\le l\le L}|y_{(k-1)L+\ell}-\bar{y}_{k}|=|y_{(k-1)L+\ell_{k}}-\bar{y}_{k}|. \ k=1,..,K.$$ 
Then over each partition $k$, we designate 
$$x_{(k-1)L+\ell_{k}}=t_{0k}, \ k=1,..,K.$$
as a potential knot, and use $x, x^2, x^3$ and $\{(x-t_{0k})_{+}^{3}\}_{k=1}^K$ as cubic spline predictors. We return
the minimum $BIC$ for each subset, and choose the model with the overall minimum $BIC$. Finally, we fit this model via least squares regression and report the model's $R_{Adj}^{2}$value. Modeling violent crime rate using cubic spline regression, we select $K+4$ parameters to account for the potential knots that are chosen by the min/max algorithm. To see the detail documentation on using this algorithm, readers are refereed to visit https://cran.r-project.org/web/packages/Kpart/index.html.

\section{Modeling Violent Crime Rates in USA}

\begin{figure}[H]
\includegraphics[width=\columnwidth]{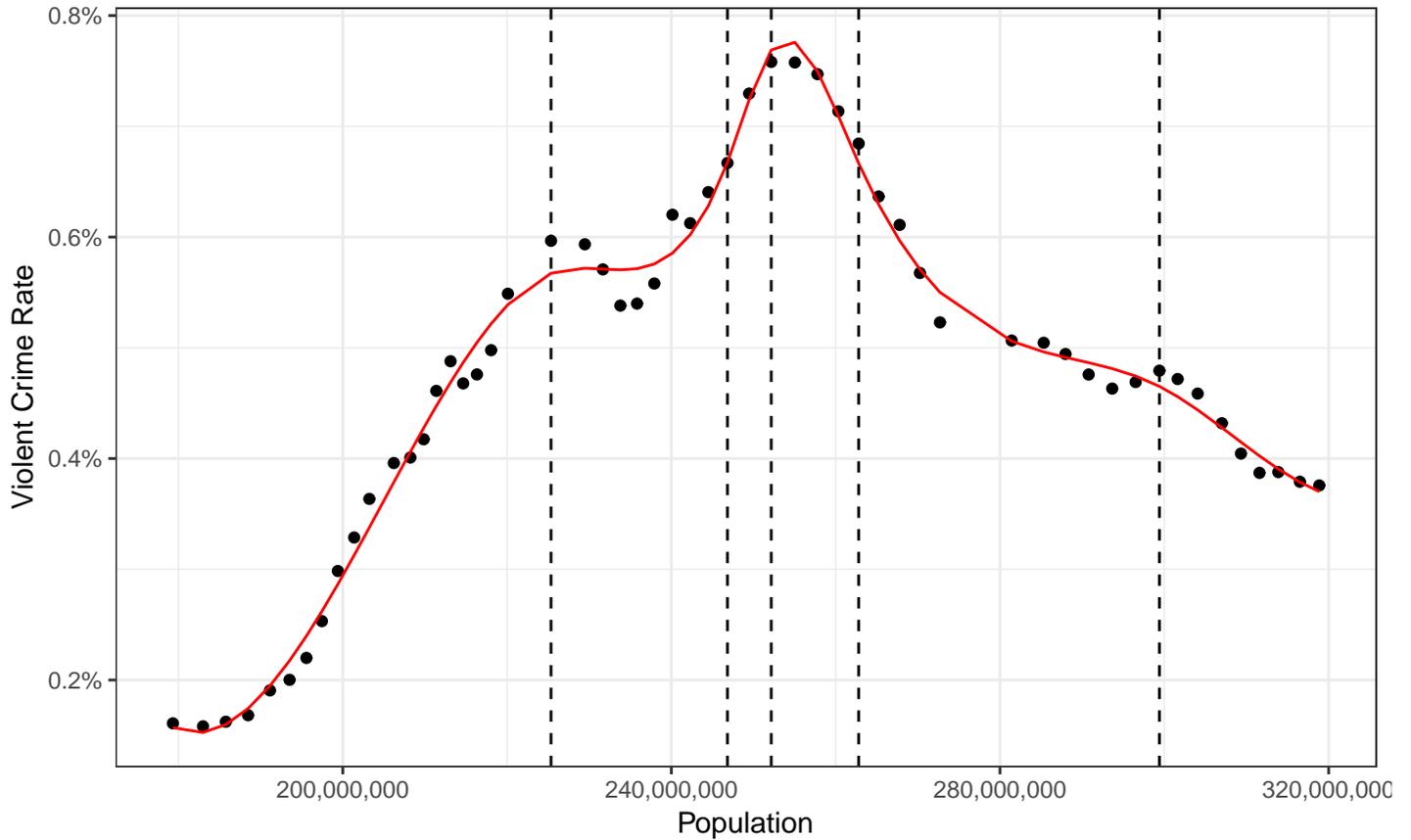}
\caption{Fitted model for USA violent crime rate}
\label{fig:usa}
\end{figure}

In this section, we illustrate the model using the min/max \textit{K-part} algorithm for entire USA and each state as well. For the entire USA, we pre-select $K$, i.e. $K=10$. This means that we partition the data exactly at census years, therefore we may have a tendency to select knots that are sequentially within a few years of each other. However,  the methodology is flexible in that a user may define any number of partitions. Within {\it Kpart} package in {\it R}, there is a function {\it part()}. The first argument of {\it part} is  {\it data} which contains a column of the average USA violent crime rate over years (USAvcr) and another column is the USA population (USApop) upon which we want to base knots. Once the data is retrieved to R, the following R code produces Figure \ref{fig:usa}. 

\texttt{> library(Kpart)}

\texttt{> fit<-part(d=data, outcomeVariable="USAvcr", splineTerm="USApop", K=10)}

\texttt{> plot(data\$USApop, data\$USvcr)}

\texttt{> lines(data\$USApop, fit\$fits) }

Figure~\ref{fig:usa} shows the overall trend for all states in USA and the fitted curve of the violent crime rate against population size. The vertical lines indicate at which population levels the knots where chosen, these values correspond to the years:  1980, 1989, 1991, 1995, and 2006. We note that this model has an $R_{Adj}^{2}=98.89\%$ 

\subsection{Fitting Florida Violent Crime Rate}
\begin{figure}[H]
\includegraphics[width=\columnwidth]{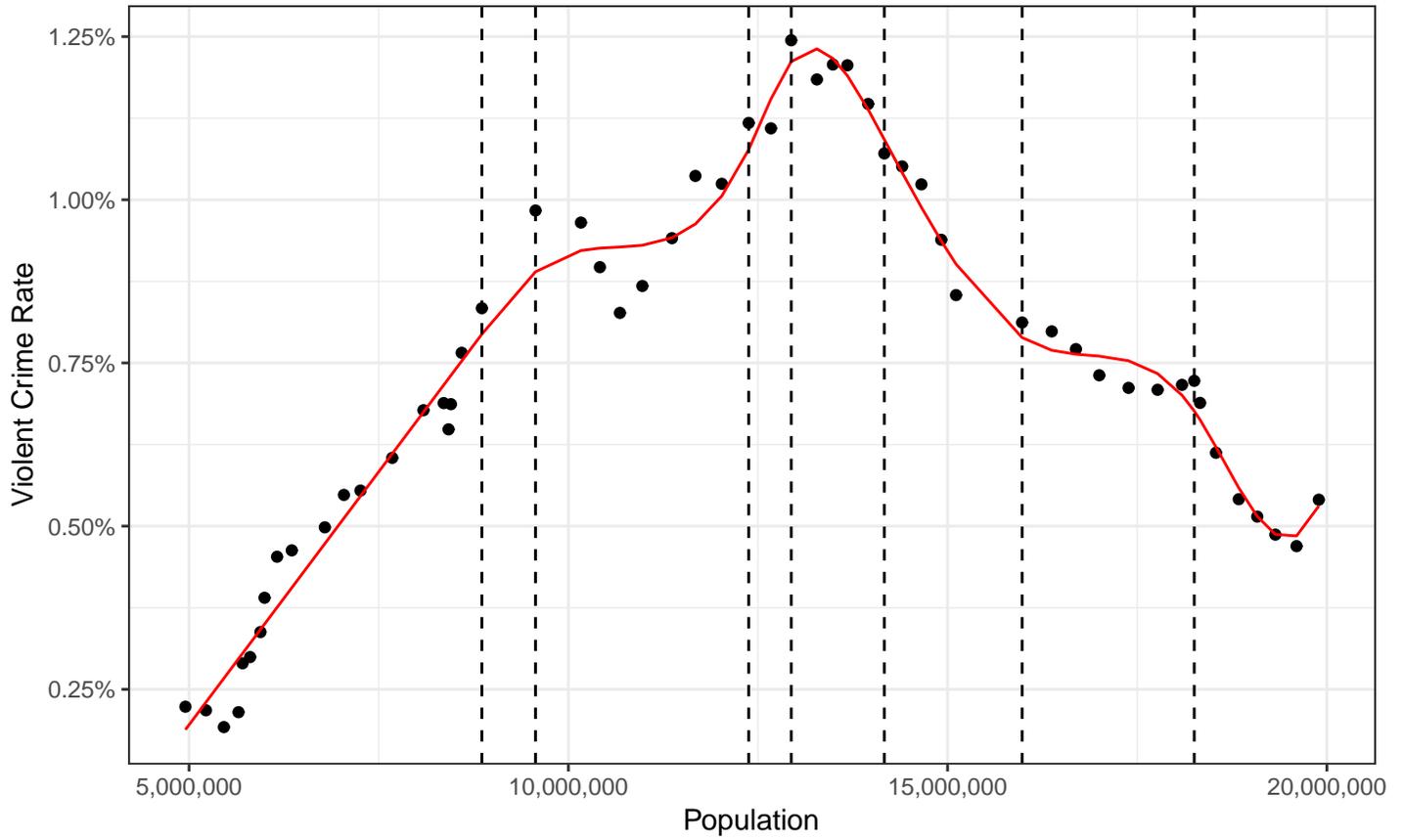}
\caption{\textit{K-part} model of Florida violent crime rate}
\label{fig:fl-dash}
\end{figure}

Represented in Florida is the general population trend of USA that the population is increasing  over time. Note that the jumps in population occur at actualized counts of the census years. The estimates that were used for intercensal population values, actually have the tendency to underestimate the actual population during these years. This causes some potential problems for knot selection. The Florida model which mirrors that of the overall USA model has an $R_{Adj}^{2}=97.67\%$. The min/max \textit{K-part} algorithm detected 7 knots as illustrated as dotted lines in Figure~\ref{fig:fl-dash}.

\subsection{Fitting Louisiana Violent Crime Rate}

\begin{figure}[H]
\includegraphics[width=\columnwidth]{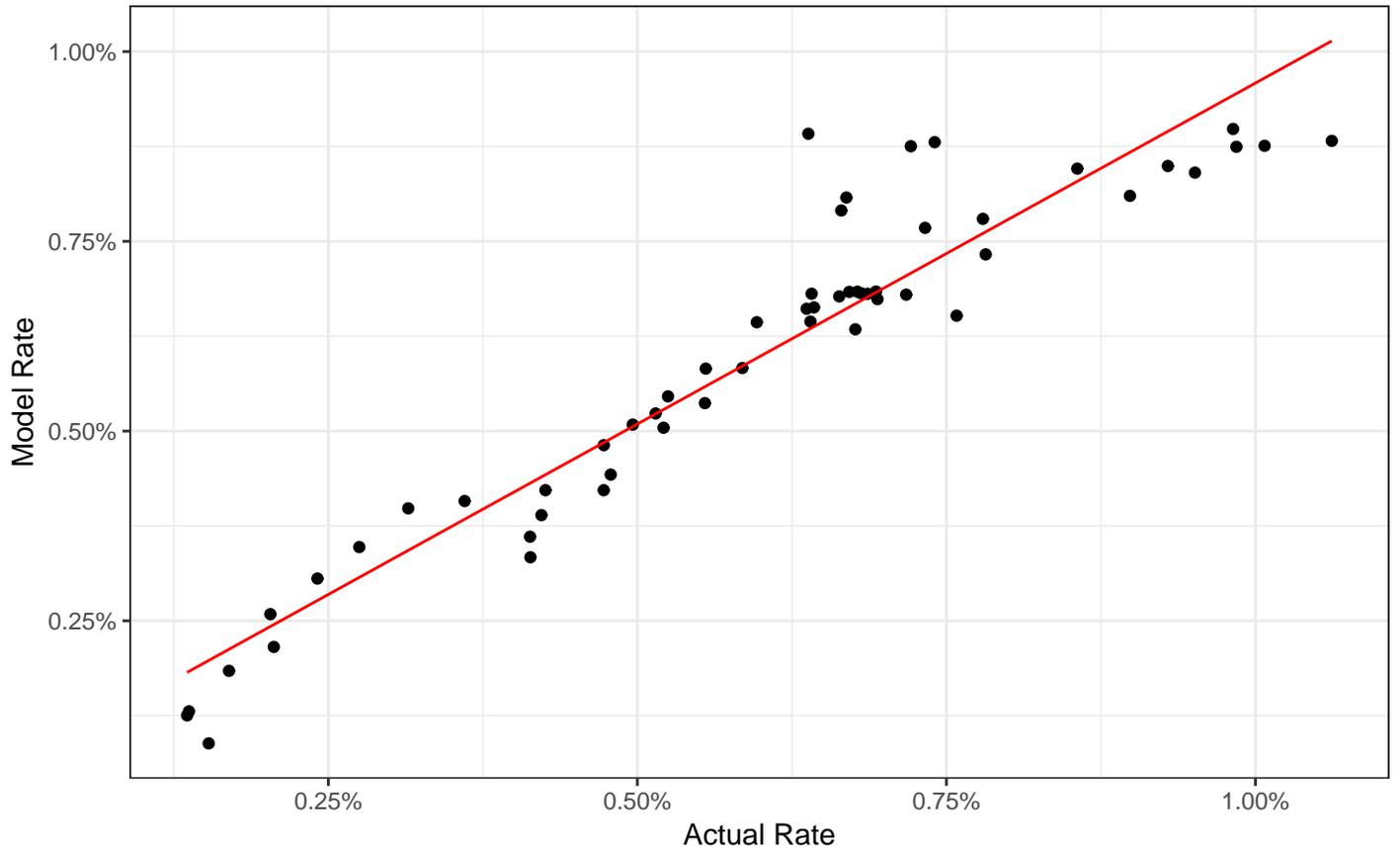}
\caption{Actual violent crime rate vs. \textit{K-part} model of Louisiana violent crime}
\label{fig:la-dash}
\end{figure}

Louisiana's trend is certainly not linear as shown in Figure~\ref{fig:la_eda}. There is  a slight deviation from population increase over time as well as notable dips close to 2005 when hurricane Katrina hit the state.  However, the min/max \textit{K-part} algorithm is still able to reasonably fit the data with an $R_{adj}^{2}=88.81\%$ using 5 knots as shown in Figure~\ref{fig:la-dash} which shows  the goodness of fit of the model to the true violent crime rate.

\subsection{Fitting Violent Crime Rate in Washington D.C.}

\begin{figure}[H]
\includegraphics[width=\columnwidth]{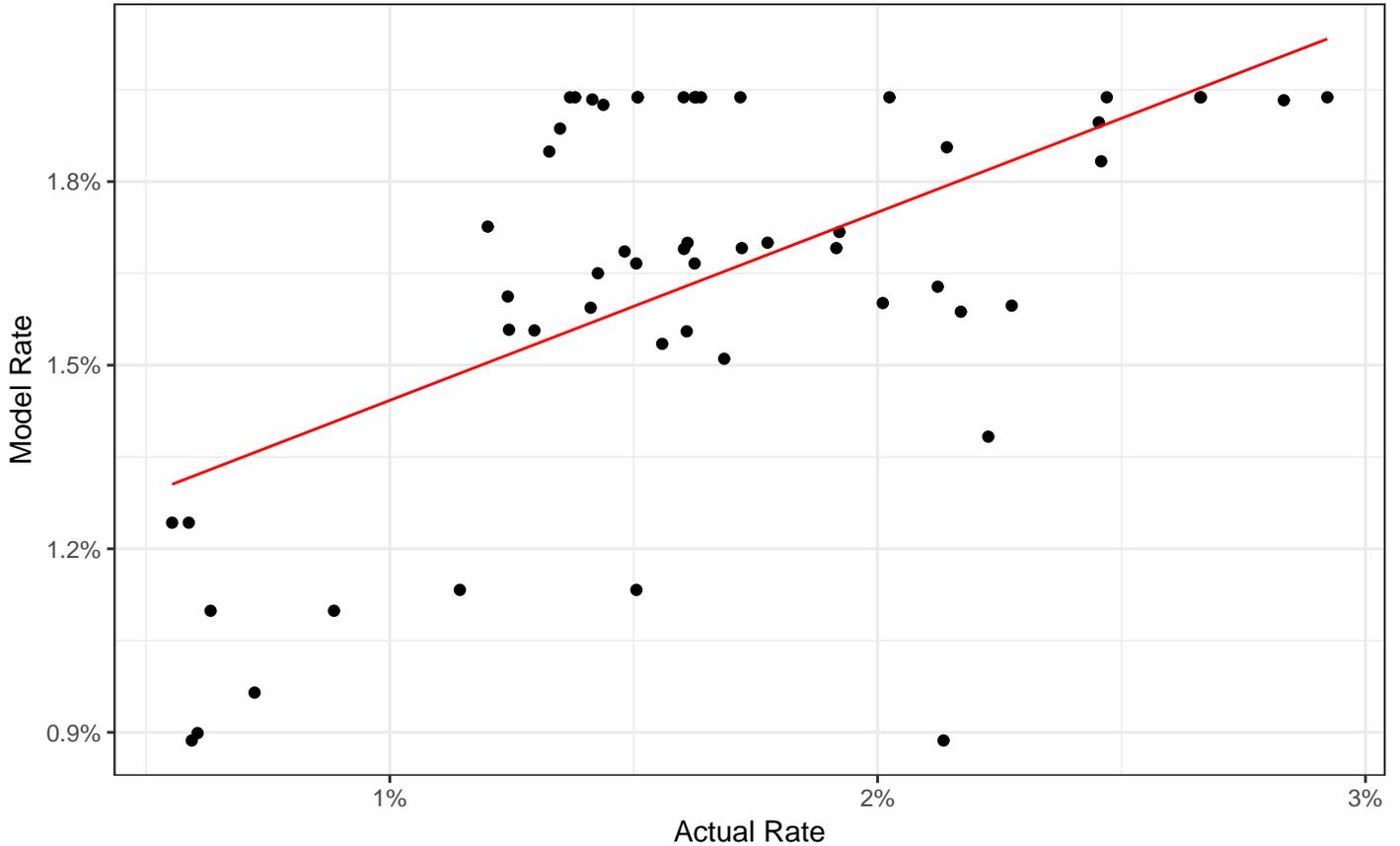}
\caption{The \textit{K-part} fit vs. actual violent crime rate of Washington DC violent crime}
\label{fig:dc-dash}
\end{figure}

For the District of Columbia, we notice that the population does not follow a linear trend over years. In this very extreme case, we see the fluctuation of population graphically by noting the many increases and decreases over time. This area shows the difficulty of our initial modeling question. We pre-select $K=5$ to simplify the model selection.  Similar to the Louisiana model, we illustrate the model's performance of the actual violent crime rate values versus those generated from the model in Figure \ref{fig:dc-dash}.

\section{Conclusions}
Upon collecting data for all fifty states, we shape a reasonable model for one state based upon it's initial scatter plot. Furthermore, an overall model was found based on the cubic spline regression model with an automated knot selection procedure. We shape an algorithm in which we can choose knots in a pre-defined number \textit{K }partitions. Upon finding these potential knots, we subset potential predictors, then find the potential model with the lowest \textit{BIC}, and finally a model is found in which all our model estimates are significant. This process can be used to model a data set for cubic spline or polynomial fitting. The R package \textit{Kpart} in
Golinko (2017) 
was used to generate the models as an extension of the thesis by Golinko (2012). 
The detailed R code for this paper can be requested from the authors.

Overall, our model with the automatic knots selection algorithm fits well for most of individual states as well as the entire USA with exception of  New York and Washington D.C. as states in which our model did not fit well. For $K=10$, the min/max \textit{K-part} knots selection algorithm for a given partition contains 5 values for entire USA violent crime rate. This corresponds to intercensal population estimates, as well as actual census year populations. The population estimates between census years tend to be underestimated, and our knot selection method reflects this phenomena. Being that there is a jump in values close to census years, the min/max algorithm commonly picks up these locations to be values of potential knots. In states where the population has a fluctuating or decreasing trend, we show caution of selecting knots based only on the relationship of violent crime and population. This was evidenced in the $R_{adj}^{2}$ value being lower than states where the population was increasing over time.

Out of the 50 states and Washington D.C., the min/max algorithm found predominantly well fitting models for all 51 cases. Specifically,  the model fits 42 states with $R_{Adj}^{2}>90\%$, and only two states below $70\%$: NY with  $R_{Adj}^{2}= 64.8\%$ and Washington D.C. with $R_{Adj}^{2}=28.10\%$. We also model the entire United States, yielding an $R_{Adj}^{2}$ of 98.89\%. The model selection process is purposefully partitioned such that we would be able to reflect census years as well as mid-census years. For the general $K$-partition case, the choice of \textit{K} is more data specific.

The results offer a simple solution to an even more general problem  researched by 
Olson (1990) and Robinson et al. (1993).
The problem focuses on how many knots are ideal, and how may we best identify these values, ultimately with the goal to minimize the number of parameters to choose. The use of computer technology makes this problem more accessible as we can simulate many cases and potential models. An extension of the idea we present here is to optimize the selection of the number \textit{K} for any given set of data.

\end{document}